\documentclass[twocolumn,showpacs,showkeys,preprintnumbers,amsmath,amssymb]{revtex4}

\usepackage{graphicx}

\begin{document}


\title{Analytical Study on the Sunyaev-Zeldovich Effect for Clusters of Galaxies. II. comparison of covariant formalisms}
\author{Satoshi Nozawa}
 \email{snozawa@josai.ac.jp}
\affiliation{
Josai Junior College, 1-1 Keyakidai, Sakado-shi, Saitama, 350-0295,
Japan}

\author{Yasuharu Kohyama and Naoki Itoh}
\affiliation{
Department of Physics, Sophia University, 7-1 Kioi-cho, Chiyoda-ku,
Tokyo, 102-8554, Japan}

\date{\today}

\begin{abstract}
  We study a covariant formalism for the Sunyaev-Zeldovich effects developed in the previous papers by the present authors, and derive analytic expressions for the redistribution functions in the Thomson approximation.  We also explore another covariant formalism recently developed by Poutanen and Vurm.  We show that the two formalisms are mathematically equivalent in the Thomson approximation which is fully valid for the cosmic microwave background photon energies.  The present finding will establish a theoretical foundation for the analysis of the Sunyaev-Zeldovich effects for the clusters of galaxies.
\end{abstract}



\maketitle

\section{Introduction}

  The Sunyaev-Zeldovich (SZ) effect\cite{suny72}, which arises from the Compton scattering of the cosmic microwave background (CMB) photons by hot electrons in clusters of galaxies (CG), provides a useful method for studies of cosmology.  For the reviews, for example, see Birkinshaw\cite{birk99} and Carlstrom, Holder, and Reese\cite{carl02}.  The original SZ formula has been derived from the Kompaneets equation\cite{komp57} in the nonrelativistic approximation.  However, recent x-ray observations (for example, Schmidt et al.\cite{tuck98} and Allen et al.\cite{alle02}) have revealed the existence of high-temperature CG such as $k_{B} T_{e} \simeq $20keV.  For such high-temperature CG, the relativistic corrections to the SZ effect will become extremely important.

  On the other hand, it has been known theoretically for some time that the relativistic corrections become significant at the short wave length region $\lambda <$ 1mm.  In particular, the recent report\cite{zemc10} on the first detection of the SZ effect at $\lambda <$ 650$\mu$m by {\it Herschel Survey} seems to confirm the relativistic corrections\cite{noza00}.  Therefore, reliable theoretical studies on the relativistic SZ effect at short wave length region will become extremely important for both existing and future observation projects.

  The relativistic SZ effect has been studied theoretically in several different approaches.  Wright\cite{wrig79} and Rephaeli\cite{reph95} calculated the photon frequency redistribution function in the electron rest frame using the scattering probability derived by Chandrasekhar\cite{chan50}, which is called as the radiative transfer method.  The second approach is the relativistic generalization of the Kompaneets equation\cite{komp57}, where the relativistically covariant Boltzmann collisional equation is solved for the photon distribution function.  This approach was used by Challinor and Lasenby\cite{chal98} and Itoh, Kohyama, and Nozawa\cite{itoh98}, which is called the covariant formalism.  Although the two are very different approaches, the obtained results for the SZ effect agreed extremely well.  This has been a longstanding puzzle in the field of the relativistic study of the SZ effect for the last ten years.  Very recently, however, Nozawa and Kohyama\cite{noza09a} showed that the two formalisms were indeed mathematically equivalent in the Thomson approximation.  This explained the reason why the two different approaches produced same results for the SZ effect even in the relativistic energies for electrons.

  On the other hand, there is yet another covariant formalism which also starts with the relativistic kinetic equation for photons\cite{nagi93, pout10}.  Although the starting equations are same for the two covariant formalisms, the final expressions for the SZ effect differ significantly from each other.  In the present paper, we explore the two covariant formalisms for the thermal SZ effect in detail.  We will show that the two formalisms are indeed mathematically equivalent in the Thomson limit, which is fully valid for the CMB photon energies.  It is also found that the two formalisms give the same expression for the kinematical SZ effect in the Thomson limit.  This will conclude that the existing formalisms\cite{wrig79,noza09a,pout10} for the SZ effects for the CG are equivalent in the Thomson approximation.  The present finding will establish a theoretical foundation for the analysis of the SZ effects for the CG.

  The present paper is organized as follows:  In Sec.~II, we derive analytic expressions of the redistribution functions for the SZ effects with the covariant formalism of Nozawa, Kohyama, and Itoh\cite{noza09a, noza09b}.  In Sec.~III, we then rewrite the redistribution functions of Poutanen and Vurm\cite{pout10} formalism in the Thomson approximation.  We will show that the redistribution functions in the two formalisms are identical.  The isotropic photon scattering approximation is also presented.  Finally, concluding remarks are given in Sec.~IV.

\section{Covariant formalism of Nozawa, Kohyama, and Itoh}

\subsection{Rate equations in the Thomson approximation}

  Recently, the present authors have published a series of papers\cite{noza09a,noza09b,noza10a,noza10b} on the Compton scattering of the CMB photons.  In the present subsection, we summarize the formalism (denoted NKI formalism hereafter) for the SZ effects of the CG.

  In Nozawa and Kohyama\cite{noza09a}, it was shown that the covariant formalism\cite{itoh98} and radiative transfer method\cite{wrig79} were mathematically equivalent in the following (Thomson) approximation:
\begin{eqnarray}
&&\hspace{-10mm}
\gamma \frac{\omega}{m} \ll 1  \, ,
\label{eq2a-1}   \\
&&\hspace{-10mm}
\gamma = \frac{1}{\sqrt{1-\beta^{2}}}  \, ,
\label{eq2a-2}
\end{eqnarray}
where $\omega$ is the photon energy, $\gamma$ is the Lorentz factor, and $\beta$ and $m$ are the velocity and rest mass of the electron, respectively.  Throughout this paper, we use the natural unit $\hbar = c = 1$, unless otherwise stated explicitly.  For the CMB photons, Eq.~(\ref{eq2a-1}) is fully valid from nonrelativistic electrons to extreme-relativistic electrons of the order of TeV region.  The kinematics in the present paper are defined as follows:  The CG is moving with a bulk velocity $\vec{\beta}_{C}$ (=$\vec{v}_{C}/c$) with respect to the CMB frame.  As a reference system, we choose the system that is fixed to the CMB.  The $z$ axis is fixed to a line connecting the observer and the center of mass of the CG.  (We assume that the observer is fixed to the CMB frame.)  We choose the positive direction of the $z$ axis as the direction from the observer to the CG.

  The rate equation for the photon distribution function $n(x)$ was derived by the present authors\cite{noza09a, noza09b} under the assumption of Eq.~(\ref{eq2a-1}).  Here, $x=\omega/k_{B}T_{\rm CMB}$ is the photon energy in units of the thermal energy of the CMB, and $s$ is the frequency shift defined by $e^{s}=x^{\prime}/x$.  We recall the results here to make the present paper more self-contained.  They are given as follows\cite{noza09a, noza09b}:
\begin{eqnarray}
&&\hspace{-10mm}
\frac{\partial n(x)}{\partial \tau}
 = \int_{-\infty}^{\infty}ds P_1(s,\beta_{C,z})
\left[n(e^sx)- n(x)\right] \, ,
\label{eq2a-3}   \\
&&\hspace{0mm}
P_1(s,\beta_{c,z}) = P_{1}(s) + \beta_{C,z} P_{\rm 1, K}(s)  \, ,
\label{eq2a-4} \\
&&\hspace{+12mm}
d\tau  =  n_e\sigma_T dt \, , 
\label{eq2a-5}
\end{eqnarray}
where $\beta_{C,z}$ is the bulk velocity of the CG parallel to the $z$-axis, $n_{e}$ is the electron number density, and $\sigma_{T}$ is the Thomson scattering cross section.  It should be noted that $O(\beta_{C}^{2})$ and higher-order contributions were neglected in deriving Eq.~(\ref{eq2a-4}), because $\beta_{C} \ll 1$ is satisfied for most of the CG.

  In Eq.~(\ref{eq2a-4}), $P_{1}(s)$ is the frequency redistribution function, and $P_{1,\rm K}(s)$ is the term which appears in the case of non-zero bulk motions.  They are defined by
\begin{eqnarray}
&&\hspace{-11mm}
P_1(s) = \int_{\beta_{min}}^{1}d\beta\beta^2\gamma^5 p_e(\gamma)P(s,\beta)   \, ,
\label{eq2a-6}  \\
&&\hspace{-11mm}
 P_{\rm 1,K}(s) = \int_{\beta_{min}}^{1}d\beta\beta^2\gamma^5 p_e(\gamma)P_{\rm K}(s,\beta)  \, ,
\label{eq2a-7}
\end{eqnarray}
where
\begin{eqnarray}
&&\hspace{-11mm}
P(s,\beta) = \frac{e^{s}}{2\beta\gamma^4}
\int_{\mu_1(s)}^{\mu_2(s)}d\mu_0 \frac{1}{(1-\beta\mu_0)^2} f\left(\mu_0, \mu_0^{\prime} \right)   \, ,
\label{eq2a-8}  \\
&&\hspace{-10mm}
P_{\rm K}(s,\beta) = \delta(\beta) \left[ \tilde{P}_{\rm K}(s, \beta) - P(s, \beta) \right]  \, ,
\label{eq2a-9}  \\
&&\hspace{-10mm}
\tilde{P}_{\rm K}(s,\beta)
= \frac{e^{s}}{2\beta\gamma^6}
\int_{\mu_1(s)}^{\mu_2(s)}d\mu_0
\frac{1}{(1-\beta\mu_0)^3}
f\left(\mu_0, \mu_{0}^{\prime} \right)   \, ,
\label{eq2a-10}  \\
&&\hspace{-11mm}
f(\mu_0,\mu_0^{\prime}) = \frac{3}{8}\left[
  1 + \mu_0^2\mu_0^{\prime 2}+\frac{1}{2}(1-\mu_0^2)(1-\mu_0^{\prime 2})
\right]  \, .
\label{eq2a-11}
\end{eqnarray}
In Eq.~(\ref{eq2a-9}), $\delta(\beta)$ is a factor related to the electron distribution function, which is, in general, a function of $\beta$.  The explicit forms are given by Nozawa and Kohyama\cite{noza09a} for three different electron distribution functions.  In the present paper, for simplicity, we use the thermal electron distribution for $p_{e}(\gamma)$, which implies
\begin{eqnarray}
&&\hspace{-10mm}
\delta( \beta ) = \frac{\gamma}{\theta_{e}}  \, ,
\label{eq2a-12}
\end{eqnarray}
where $\theta_{e}=k_{B} T_{e}/mc^{2}$ is the electron thermal energy of the CG in units of the electron rest energy.  The electron thermal distribution function is
\begin{eqnarray}
&&\hspace{-10mm}
p_{e}(\gamma) = \frac{1}{\theta_{e} K_{2}(1/\theta_{e})} \exp(-\gamma /\theta_{e}) \, ,
\label{eq2a-13}
\end{eqnarray}
where $K_{2}(z)$ is the modified Bessel function of the second kind.  Variables appearing in Eqs.~(\ref{eq2a-6}) -- (\ref{eq2a-11}) are summarized as follows:
\begin{eqnarray}
&&\hspace{-10mm}
\beta_{min} = (1-e^{-|s|})/(1+e^{-|s|})  \, ,
\label{eq2a-14} \\
&&\hspace{-10mm}
\mu_{0}^{\prime} = [1-e^s(1-\beta\mu_0)]/\beta  \, ,
\label{eq2a-15}  \\
&&\hspace{-10mm}
\mu_1(s) = \left\{
\begin{array}{ll}
-1 &\quad  {\rm for} \, \, \, s \leq 0 \\
{[1-e^{-s}(1+\beta)]/\beta} &\quad {\rm for} \, \, \, s > 0
\end{array}
\right.  \, ,
\label{eq2a-16} \\
&&\hspace{-10mm}
\mu_2(s) = \left\{
\begin{array}{ll}
{[1-e^{-s}(1-\beta)]/\beta} &\quad {\rm for} \, \, \, s < 0 \\
1 &\quad  {\rm for} \, \, \, s \geq 0 
\end{array}
\right. \, .
\label{eq2a-17}
\end{eqnarray}

  For most of CG , $\tau \ll 1$ is satisfied.  Then one obtains the following solution for Eq.~(\ref{eq2a-3}):
\begin{eqnarray}
&&\hspace{-10mm}
\Delta n(x) = n(x) - n_{0}(x)
\nonumber    \\
&&\hspace{1mm}
\equiv  \Delta n_{t}(x) + \beta_{C,z} \Delta n_{k}(x)  \, ,
\label{eq2a-18}
\end{eqnarray}
\begin{eqnarray}
&&\hspace{-10mm}
\Delta n_{t}(x) =  \tau \left[\int_{-\infty}^{\infty}ds P_1(s) n_{0}(e^sx)- n_{0}(x) \right]  \, ,
\label{eq2a-19} \\
&&\hspace{-10mm}
\Delta n_{k}(x) =  \tau \left[\int_{-\infty}^{\infty}ds P_{\rm 1,K}(s) n_{0}(e^sx)- n_{0}(x) \right] \, ,
\label{eq2a-20}
\end{eqnarray}
where
\begin{eqnarray}
&&\hspace{-10mm}
n_{0}(x) = \frac{1}{e^{x}-1}
\label{eq2a-21}
\end{eqnarray}
is the initial Planckian photon distribution function.  Equations (\ref{eq2a-19}) and (\ref{eq2a-20}) correspond to the thermal SZ effect and kinematical SZ effect, respectively.

\subsection{Analytic expressions for $P(s, \beta)$ and $P_{\rm K}(s, \beta)$}

  First, we show the analytic expression for the redistribution function $P(s, \beta)$ in the NKI formalism.  In Eq.~(\ref{eq2a-8}), the integral of $\mu_{0}$ can be done analytically.  The explicit form was derived by En$\ss$lin and Biermann\cite{enss98} and also given by Eqs.~(18) and (19) in Ref.\cite{noza10a}.  In the present paper, we rewrite the expression into the following form:
\begin{eqnarray}
&&\hspace{-10mm}
P(s,\beta) = e^{3s/2} Q(s, \beta)  \, ,
\label{eq2b-1}  \\
&&\hspace{-10mm}
Q(s,\beta) =\frac{3}{16\beta^2\gamma^4}
\biggl[
\nonumber  \\
&&\hspace{8mm}
\frac{4}{\beta^3}\left\{ 3+\gamma^2\beta^4+\frac{\beta^2-3}{2\beta}
\ln\frac{1+\beta}{1-\beta}\right\} \cosh\frac{s}{2}
\nonumber \\
&&\hspace{6mm}
-\frac{1}{\beta^4}\left\{ \frac{1}{\gamma^2}\sinh\frac{3|s|}{2}
+{2}(\beta^2-3)|s|\cosh\frac{s}{2}
\right.
\nonumber \\
&&\hspace{16mm}
\left.\left.
+
\left(9-\beta^2+4\gamma^2\beta^4\right)\sinh\frac{|s|}{2}
\right\}
\right]  \, .
\label{eq2b-1b}
\end{eqnarray}
Note that $Q(s,\beta)$ is an even function on $s$, namely $Q(s,\beta)=Q(-s,\beta)$.  Thus, the non-symmetric structure of the redistribution function $P_{1}(s)$ is due to the function $e^{3s/2}$ in Eq.~(\ref{eq2b-1}), because $\beta_{min}$ in Eq.~(\ref{eq2a-6}) is also an even function on $s$.

  Similarly, we show the analytic expression for the redistribution function $P_{\rm K}(s, \beta)$ in the NKI formalism.  The explicit form for $\tilde{P}_{\rm K}(s, \beta)$ was given by Eqs.~(25) and (26) in Ref.\cite{noza10b}.  In the present paper, we rewrite the expression into the following form:
\begin{eqnarray}
&&\hspace{-5mm}
\tilde{P}_{\rm K}(s,\beta) = e^{2s} \tilde{Q}_{\rm K}(s,\beta)  \, ,
\label{eq2b-2}  \\
&&\hspace{-5mm}
\tilde{Q}_{\rm K}(s,\beta) = \frac{3}{16\beta^2\gamma^4}
\biggl[
\nonumber \\
&&\hspace{3mm}
\frac{4}{\beta^3}\left\{ -3+2\beta^2+\frac{3}{2\beta\gamma^2}
\ln\frac{1+\beta}{1-\beta}
\right\}
\nonumber \\
&&\hspace{0mm}
+\frac{2}{\beta^3}\left\{ -3+3\beta^2+2\gamma^2\beta^4
+\frac{3-\beta^2}{2\beta\gamma^2}
\ln\frac{1+\beta}{1-\beta}
\right\} \cosh{s}
\nonumber \\
&&\hspace{0mm}
-\frac{1}{\beta^4}\left\{ \frac{6}{\gamma^2}|s|
+\frac{3-\beta^2}{\gamma^2}|s|\cosh{s}
\right.
\nonumber \\
&&\hspace{10mm}
- \left(9-10\beta^2-\beta^4-4\gamma^2\beta^6\right)\sinh{|s|}
\biggr\}
\biggr]  \, .
\label{eq2b-2b}
\end{eqnarray}
Note that $\tilde{Q}_{\rm K}(s,\beta)$ is also an even function on $s$.  Inserting Eqs.~(\ref{eq2b-1}) and (\ref{eq2b-2}) into Eq.~(\ref{eq2a-9}), one obtains the analylic expression for $P_{\rm K}(s, \beta)$.  Thus, the full analytic expressions for $P(s, \beta)$ and $P_{\rm K}(s, \beta)$ have been derived in the NKI formalism.

  Before closing the present section, it is worth to mention the following.  The thermal SZ effect (Eq.~(\ref{eq2a-19})) and kinematical SZ effect (Eq.~(\ref{eq2a-20})) are obtained by the double integrals over the variables $s$ and $\beta$, because the analytic forms for $P(s, \beta)$ and $P_{\rm K}(s, \beta)$ are given in terms of Eqs.~(\ref{eq2b-1}) and (\ref{eq2b-2}).  This makes the numerical calculations of the SZ effects extremely fast, which will be quite useful for the analysis of the observation data.  The numerical programs are available upon request from one of the present authors (S. N.).

\section{Equvalence between two formalisms in the Thomson approximation}

\subsection{Thermal SZ effect in the Thomson approximation}

  In the present subsection, we show that the formalism shown by Poutanen and Vurm\cite{pout10} (denoted PV hereafter) for the thermal SZ effect ($\beta_{C}=0$ case) is equivalent in the Thomson approximation to the NKI formalism.  Before to proceed the calculation, it should be remarked the following.  In the present paper, the photon energies are expressed in units of the thermal energy of the CMB.  On the other hand, all energy variables in the PV paper are in units of the electron rest energy.  In order to make the present paper self-consistent, we introduce new variables $x_{\rm PV}$ and $x'_{\rm PV}$ which correspond to $x$ and $x_{1}$ in the PV paper as follows:
\begin{eqnarray}
&&\hspace{-10mm}
x_{\rm PV} = \frac{\omega}{mc^2} = x \theta_{\rm CMB}  \, ,
\label{eq3a-1}  \\
&&\hspace{-10mm}
x'_{\rm PV} = \frac{\omega'}{mc^2} = x' \theta_{\rm CMB}  \, ,
\label{eq3a-2}
\end{eqnarray}
where $\theta_{\rm CMB} = k_{B} T_{\rm CMB}/mc^2$.

  The source function is defined by Eq.~(174) of the PV paper.  We rewrite the expression with variables defined in the present paper as follows:
\begin{eqnarray}
&&\hspace{-10mm}
S_{\rm PV}(x) = \tau \int_{-\infty}^{\infty}ds\, P_{1, \rm PV}(s) n_{0}(e^sx)  \, ,
\label{eq3a-3}  \\
&&\hspace{-10mm}
P_{1, \rm PV}(s) = \int_{\beta_{\rm{min}}}^{1}d\beta\beta^2\gamma^5p_e(\gamma) P_{\rm PV}(s, \beta)  \, ,
\label{eq3a-4}  \\
&&\hspace{-10mm}
P_{\rm PV}(s, \beta) = \frac{3}{16\beta\gamma^2} x_{\rm PV} e^{2s} \int_{\mu_{\rm{min}}}^{\mu_{\rm{max}}} d\mu R_0  \, ,
\label{eq3a-5}
\end{eqnarray}
where $\mu_{\rm min}$ and $\mu_{\rm max}$ are the minimum and maximum values of the cosine of the scattering angle, and the function $R_{0}$ is given by Eq.~(E2) of the PV paper.  It should be noted that the angular integrals in Eq.~(\ref{eq3a-5}) were performed by Brinkmann\cite{brin84} and also by Nagirner and Poutanen\cite{nagi94}.  The source function is related to the change of the photon occupation number by
\begin{eqnarray}
&&\hspace{-10mm}
\Delta n_{t, \rm PV}(x) = S_{\rm PV}(x) - \tau n_{0}(x)  \, ,
\label{eq3a-6}
\end{eqnarray}
which corresponds to Eq.~(\ref{eq2a-19}).

  In the Thomson approximation ($x'_{\rm PV} \ll 1$), one finds
\begin{eqnarray}
&&\hspace{-10mm}
\mu_{\rm min} = -1  \, ,
\label{eq3a-7}  \\
&&\hspace{-10mm}
\mu_{\rm max} = 1 - \frac{e^{-s}(e^s-1)^2}{2\beta^2\gamma^2}  \, .
\label{eq3a-8}
\end{eqnarray}
The leading-order terms (1/$x_{\rm PV}$ terms) of $R_{0}$ can be rewritten in the Thomson approximation by
\begin{eqnarray}
&&\hspace{-10mm}
R_0 \approx \frac{2}{Q}-\frac{2}{q}\frac{d}{a^3}+\frac{d}{2q^2a^5}
\left(5\frac{d^2}{a^2}-3Q^2\right) \, .
\label{eq3a-9}
\end{eqnarray}
Note that $a_{\pm} \approx a$ and $d_{\pm} \approx d$ were used in deriving Eq.~(\ref{eq3a-9}).  The variables appearing in Eq.~(\ref{eq3a-9}) are expressed by:
\begin{eqnarray}
&&\hspace{-20mm}
Q = x_{\rm PV}\sqrt{1+e^{2s}-2e^s\mu}  \, ,
\label{eq3a-10} \\
&&\hspace{-20mm}
d = x_{\rm PV}\gamma (e^s+1)  \, ,
\label{eq3a-11} \\
&&\hspace{-20mm}
a = \sqrt{\gamma^2+r}  \, ,
\label{eq3a-12}  \\
&&\hspace{-20mm}
q = x_{\rm PV}^2e^s(1-\mu)  \, ,
\label{eq3a-13} \\
&&\hspace{-20mm}
r = \frac{1+\mu}{1-\mu}  \, .
\label{eq3a-14}
\end{eqnarray}

  Inserting Eq.~(\ref{eq3a-9}) into Eq.~(\ref{eq3a-5}) and expressing in terms of the variables of Eqs.~(\ref{eq3a-10})--(\ref{eq3a-14}), one can write the redistribution function as follows:
\begin{eqnarray}
&&\hspace{-7mm}
P_{\rm{PV}}(s,\beta)
=\frac{3}{32\beta^2\gamma^4} \Bigl[
4\beta\gamma^2e^{2s}I_1
-4\beta\gamma^3 e^s(e^s+1)I_2
\nonumber  \\
&&\hspace{13mm}
+\beta\gamma^5(e^s+1)^3I_3
-\beta\gamma^3(e^s-1)^2(e^s+1)I_4
\nonumber  \\
&&\hspace{13mm}
-2\beta\gamma^3 e^s(e^s+1)I_5
\Bigr]  \, ,
\label{eq3a-15}
\end{eqnarray}
where $I_{1}, \dots, I_{5}$ are the integrals defined by
\begin{eqnarray}
&&\hspace{-20mm}
I_1 =
\int_{-1}^{\mu_{\rm{max}}}d\mu
\frac{1}{(1+e^{2s}-2e^s\mu)^{1/2}}  \, ,
\label{eq3a-16} \\
&&\hspace{-20mm}
I_2 =
\int_{-1}^{\mu_{\rm{max}}}d\mu
\frac{1}{(1-\mu)(\gamma^2+r)^{3/2}}  \, ,
\label{eq3a-17} \\
&&\hspace{-20mm}
I_3 =
5\int_{-1}^{\mu_{\rm{max}}}d\mu
\frac{1}{(1-\mu)^2(\gamma^2+r)^{7/2}}  \, ,
\label{eq3a-18} \\
&&\hspace{-20mm}
I_4 =
3\int_{-1}^{\mu_{\rm{max}}}d\mu
\frac{1}{(1-\mu)^2(\gamma^2+r)^{5/2}}  \, ,
\label{eq3a-19} \\
&&\hspace{-20mm}
I_5 =
3\int_{-1}^{\mu_{\rm{max}}}d\mu
\frac{1}{(1-\mu)(\gamma^2+r)^{5/2}}  \, .
\label{eq3a-20}
\end{eqnarray}
Their elementary integrals can be performed.  One has as follows:
\begin{eqnarray}
&&\hspace{-10mm}
I_1 = \frac{e^{-s}}{\beta}\left[(e^s+1)\beta-|e^s-1|\right]  \, ,
\label{eq3a-21}
\end{eqnarray}
\begin{eqnarray}
&&\hspace{-10mm}
I_2 = I_{21} + I_{22}  \, ,
\label{eq3a-22} \\
&&\hspace{-10mm}
I_{21} = \frac{2}{\gamma}-\frac{1}{2\beta^3\gamma^3}\frac{e^{-s}(e^s-1)^2|e^s-1|}{(e^s+1)}  \, ,
\label{eq3a-23} \\
&&\hspace{-10mm}
I_{22} = -\frac{2}{\beta^2\gamma}
+\frac{1}{2\beta^3\gamma^3}e^{-s}(e^s+1)|e^s-1|
\nonumber  \\
&&\hspace{0mm}
+\frac{1}{\beta^3\gamma^3}(\lambda_{\beta}-|s|)  \, ,
\label{eq3a-24}
\end{eqnarray}
\begin{eqnarray}
&&\hspace{-10mm}
I_3 = \frac{1}{\gamma^5}-\frac{1}{\beta^5\gamma^5}\frac{(e^s-1)^4|e^s-1|}{(e^s+1)^5}  \, ,
\label{eq3a-25}
\end{eqnarray}
\begin{eqnarray}
&&\hspace{-10mm}
I_4 = \frac{1}{\gamma^3}-\frac{1}{\beta^3\gamma^3}\frac{(e^s-1)^2|e^s-1|}{(e^s+1)^3}  \, ,
\label{eq3a-26}
\end{eqnarray}
\begin{eqnarray}
&&\hspace{-10mm}
I_5 = I_{51} + I_{52}  \, ,
\label{eq3a-27} \\
&&\hspace{-10mm}
I_{51}= \frac{2}{\gamma^3}-\frac{1}{2\beta^5\gamma^5}\frac{e^{-s}(e^s-1)^4|e^s-1|}{(e^s+1)^3}  \, ,
\label{eq3a-28}  \\
&&\hspace{-10mm}
I_{52} = -\frac{2}{\beta^4\gamma^3}(3-2\beta^2)
+\frac{4}{\beta^5\gamma^5}\frac{|e^s-1|}{e^s+1}
\nonumber  \\
&&\hspace{0mm}
+\frac{1}{2\beta^5\gamma^5}e^{-s}(e^s+1)|e^s-1|
\nonumber  \\
&&\hspace{0mm}
+\frac{3}{\beta^5\gamma^5}(\lambda_{\beta}-|s|)
\label{eq3a-29}  \, ,
\end{eqnarray}
where
\begin{eqnarray}
&&\hspace{-20mm}
\lambda_{\beta} = \ln \frac{1+\beta}{1-\beta}  \, .
\label{eq3a-30}
\end{eqnarray}

  Inserting Eqs.~(\ref{eq3a-21})--(\ref{eq3a-29}) into Eq.~(\ref{eq3a-15}), and performing a straightforward calculation, one finds that $P_{\rm PV}(s, \beta)$ coincides with $P(s, \beta)$ given by Eqs.~(\ref{eq2b-1}) and (\ref{eq2b-1b}).  Thus, the PV formalism is equivalent to the NKI formalism in the Thomson approximation for the thermal SZ effect.  It should be remarked that $I_{3}$, $I_{4}$ and $I_{51}$ terms in Eq.~(\ref{eq3a-15}) have cancelled out completely in deriving $P_{\rm PV}(s, \beta)$.

  Before closing the present subsection, it should remarked that the equivalence of the two formalisms is valid not only for electrons in the thermal distribution but also in other distributions such as power-laws.

\subsection{Kinematical SZ effect in the Thomson approximation}

  In the present subsection, we show that the PV formalism\cite{pout10} for the kinematical SZ effect ($\beta_{C} \neq 0$ case) is also equivalent to the NKI formalism.  In the PV paper, $O(\beta_{C})$ and $O(\beta_{C}^{2})$ terms are calculated for the kinematical SZ effect.  On the other hand, as discussed in the previous section, $\beta_{C} \ll 1$ is satisfied for most of the CG.  In Nozawa, Itoh, and Kohyama\cite{noza98}, it was shown that $O(\beta_{C}^{2})$ contribution can be safely neglected.  Therefore, we restrict ourselves to the $O(\beta_{C})$ terms.

  The source function of $O(\beta_{C})$ in Eq.~(174) of the PV paper is expressed by
\begin{eqnarray}
&&\hspace{-10mm}
S_{\rm K,PV}(x) = \beta_{C,z} \tau \int_{-\infty}^{\infty}ds P_{1, \rm{K PV}}(s) n_{0}(e^sx)  \, ,
\label{eq3b-1}  \\
&&\hspace{-10mm}
P_{1, \rm{K PV}}(s) = \int_{\beta_{\rm{min}}}^{1}d\beta\beta^2\gamma^5p_{e}(\gamma)P_{\rm{K,PV}}(s,\beta) \, ,
\label{eq3b-2}  \\
&&\hspace{-10mm}
P_{\rm{K,PV}}(s,\beta) = \frac{\gamma}{\theta_e}\left[\tilde{P}_{\rm{K,PV}}(s,\beta) - P_{\rm{PV}}(s,\beta) \right]  \, ,
\label{eq3b-3} \\
&&\hspace{-10mm}
\tilde{P}_{\rm{K,PV}}(s,\beta)
=\frac{3}{32\beta\gamma^3}x_{\rm PV}^2e^{3s}\int_{\mu_{\rm{min}}}^{\mu_{\rm{max}}} d\mu (1-\mu)R_{\Sigma}  \, ,
\label{eq3b-4}
\end{eqnarray}
where $P_{\rm PV}(s,\beta)$ is given by Eq.~(\ref{eq3a-5}), and the function $R_{\Sigma}$ is given by Eq.~(E3) of the PV paper.  It should be remarked that $\beta_{C,z} = -\eta \beta_{b}$ was used in deriving Eq.~(\ref{eq3b-1}), where $\eta$ is the cosine of the polar angle of the initial photon and $\beta_{b}=\beta_{C}$.  The source function is related to the photon occupation number by
\begin{eqnarray}
&&\hspace{-10mm}
\beta_{C,z} \Delta n_{k, \rm PV}(x) = S_{\rm K,PV}(x) - \beta_{C,z} \tau n_{0}(x)   \, ,
\label{eq3b-5}
\end{eqnarray}
which corresponds to Eq.~(\ref{eq2a-20}).

  The leading-order terms (1/$x^{2}_{\rm PV}$ terms) of $R_{\Sigma}$ can be rewritten in the Thomson approximation by
\begin{eqnarray}
&&\hspace{-10mm}
R_{\Sigma} \approx \frac{4d}{Q^{3}}-\frac{4}{qa}+\frac{3d^2}{q^2a^5}
-\frac{Q^2}{q^2a^3} \, .
\label{eq3b-6}
\end{eqnarray}
  Inserting Eq.~(\ref{eq3b-6}) into Eq.~(\ref{eq3b-4}) and expressing in terms of the variables of Eqs.~(\ref{eq3a-10})--(\ref{eq3a-14}), one can write the redistribution function as follows:
\begin{eqnarray}
&&\hspace{-8mm}
\tilde{P}_{\rm K,PV}(s,\beta)
=\frac{3e^s}{32\beta^2\gamma^4} \Bigl[
4\beta\gamma^2e^{2s}(e^s+1)K_{1}
-4\beta\gamma e^sK_2
\nonumber \\
&&\hspace{23mm}
+\beta\gamma^3(e^s+1)^2K_3
-\beta\gamma(e^s-1)^2K_4
\nonumber  \\
&&\hspace{23mm}
-2\beta\gamma e^sK_5
\Bigr]  \, ,
\label{eq3b-7}
\end{eqnarray}
where $K_{1}, \dots, K_{5}$ are the integrals defined by
\begin{eqnarray}
&&\hspace{-20mm}
K_1 =
\int_{-1}^{\mu_{\rm{max}}}d\mu
\frac{1-\mu}{(1+e^{2s}-2e^s\mu)^{3/2}}  \, ,
\label{eq3b-8} \\
&&\hspace{-20mm}
K_2 =
\int_{-1}^{\mu_{\rm{max}}}d\mu
\frac{1}{(\gamma^2+r)^{1/2}}  \, ,
\label{eq3b-9} \\
&&\hspace{-20mm}
K_3 =
3\int_{-1}^{\mu_{\rm{max}}}d\mu
\frac{1}{(1-\mu)(\gamma^2+r)^{5/2}}  \, ,
\label{eq3b-10} \\
&&\hspace{-20mm}
K_4 =
\int_{-1}^{\mu_{\rm{max}}}d\mu
\frac{1}{(1-\mu)(\gamma^2+r)^{3/2}}  \, ,
\label{eq3b-11} \\
&&\hspace{-20mm}
K_5 =
\int_{-1}^{\mu_{\rm{max}}}d\mu
\frac{1}{(\gamma^2+r)^{3/2}}  \, .
\label{eq3b-12}
\end{eqnarray}

Their elementary integrals can be performed.  One has as follows:
\begin{eqnarray}
&&\hspace{-10mm}
K_1 = \frac{e^{-2s}}{2\beta\gamma^2}\left[(1-2\gamma^2)|e^s-1|
+{2\beta\gamma^2}\frac{e^{2s}+1}{e^s+1}\right]  \, ,
\label{eq3b-13}  \\
&&\hspace{-10mm}
K_2 = -I_{22}  \, ,
\label{eq3b-14}  \\
&&\hspace{-10mm}
K_3 = I_{5}  \, ,
\label{eq3b-15}  \\
&&\hspace{-10mm}
K_4 = I_{2}  \, ,
\label{eq3b-16}  \\
&&\hspace{-10mm}
K_5 = -I_{52}  \, .
\label{eq3b-17}
\end{eqnarray}
  Inserting Eqs.~(\ref{eq3b-13})--(\ref{eq3b-17}) into Eq.~(\ref{eq3b-7}), and performing a straightforward calculation, one finds that $\tilde{P}_{\rm K,PV}(s, \beta)$ coincides with $\tilde{P}_{\rm K}(s, \beta)$ given by Eqs.~(\ref{eq2b-2}) and (\ref{eq2b-2b}).  Therefore
\begin{eqnarray}
P_{\rm K}(s, \beta) = P_{\rm K,PV}(s, \beta)  \, .
\label{eq3b-18}
\end{eqnarray}
Thus, the PV formalism is also equivalent to the NKI formalism in the Thomson approximation for the kinematical SZ effect.

\subsection{Redistribution functions in the isotropic scattering approximation}

  The isotropic scattering (ISO) approximation for photons has been studied, for example, in Rephaeli\cite{reph95} for the thermal SZ effect.  In the NKI formalism, the ISO approximation can be imposed by averaging over the initial photon solid angle, namely
\begin{eqnarray}
&&\hspace{-10mm}
\frac{1}{4 \pi} \int d \Omega_{k} f(\mu_{0}, \mu_{0}^{\prime}) = \frac{1}{2} \, ,
\label{eq3c-1}
\end{eqnarray}
where $f(\mu_{0}, \mu_{0}^{\prime})$ is defined by Eq.~(\ref{eq2a-11}).  Thus, the ISO approximation in the NKI formalism is obtained by replacing
\begin{eqnarray}
&&\hspace{-10mm}
f(\mu_{0}, \mu_{0}^{\prime}) \longrightarrow \frac{1}{2} \, .
\label{eq3c-2}
\end{eqnarray}
Inserting Eq.~(\ref{eq3c-2}) into Eqs.~(\ref{eq2a-8}) and (\ref{eq2a-10}), one obtains the redistribution functions in the ISO approximation as follows:
\begin{eqnarray}
&&\hspace{-10mm}
P_{\rm iso}(s, \beta) = \frac{e^{3s/2}}{2 \beta^{2}\gamma^{2}} \left[ \beta \cosh \frac{s}{2} - \sinh\frac{|s|}{2} \right]  \, ,
\label{eq3c-3} \\
&&\hspace{-10mm}
\tilde{P}_{\rm K,iso}(s, \beta) =  \frac{e^{2s}}{4 \beta^{2}\gamma^{2}} \left[ 2\beta \cosh s - \left(1+\beta^{2}\right) \sinh |s| \right] \, ,
\label{eq3c-4}  \\
&&\hspace{-10mm}
P_{\rm K,iso}(s, \beta) = \frac{\gamma}{\theta_{e}} \left[ \tilde{P}_{\rm K,iso}(s, \beta) - P_{\rm iso}(s, \beta) \right]  \, .
\label{eq3c-5}
\end{eqnarray}

\begin{figure}
\begin{center}
\includegraphics[angle=0,width=0.50\textwidth]{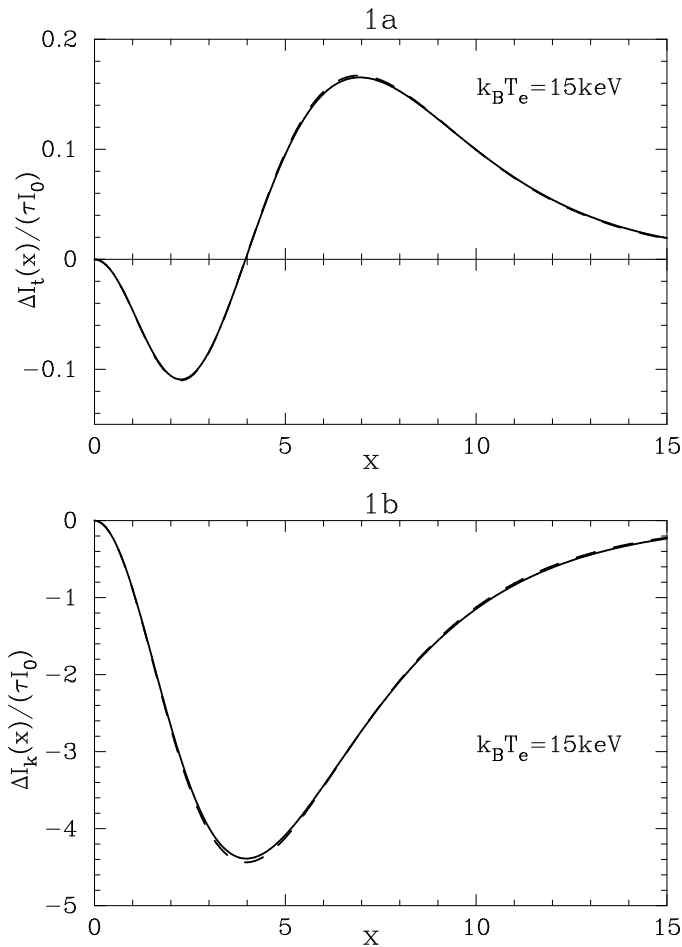}
\end{center}
\caption{Dependences of $\Delta I_{t}(x)$ and $\Delta I_{k}(x)$ for $k_{B}T_{e}$ = 15keV as a function of $x$.  Figs.~1a and 1b are $\Delta I_{t}(x)$ and $\Delta I_{k}(x)$, respectively.  The solid curve is the full calculation, and the dashed curve is the ISO approximation.}
\end{figure}

  On the other hand, the expressions of the ISO approximation in the PV formalism are given by Eqs.~(145) and (146) of the PV paper.  In the Thomson approximation, they can be rewritten by
\begin{eqnarray}
&&\hspace{-10mm}
R_{0} = \frac{4}{3Q}  \, ,
\label{eq3c-6}  \\
&&\hspace{-10mm}
R_{\Sigma}= \frac{8 d}{3 Q^{3}}  \, .
\label{eq3c-7}
\end{eqnarray}
Inserting Eqs.~(\ref{eq3c-6}) and (\ref{eq3c-7}) into Eq.~(\ref{eq3a-5}) and (\ref{eq3b-4}) and repeating the same straightforward calculations done in the previous subsections, one finally obtains the same expressions as Eqs.~(\ref{eq3c-3}) and (\ref{eq3c-4}).

  Finally, we study the accuracy of the ISO approximation compared with the full calculation.  We define the change of the spectral intensity function $\Delta I(x)$ by
\begin{eqnarray}
&&\hspace{-10mm}
\Delta I(x) = \Delta I_{t}(x) + \beta_{C,z} \Delta I_{k}(x)  \, ,
\label{eq3c-8}  \\
&&\hspace{-10mm}
\Delta I_{t}(x) = I_{0} x^{3} \Delta n_{t}(x) \, ,
\label{eq3c-9}  \\
&&\hspace{-10mm}
\Delta I_{k}(x) = I_{0} x^{3} \Delta n_{k}(x) \, ,
\label{eq3c-10}
\end{eqnarray}
where $I_{0}=(k_{B}T_{\rm CMB})^{3}/2\pi^{2}$, and $\Delta n_{t}(x)$ and $\Delta n_{k}(x)$ are given by Eqs.~(\ref{eq2a-19}) and (\ref{eq2a-20}), respectively.  

  In Fig.~1, we plot $\Delta I_{t}(x)$ and $\Delta I_{k}(x)$ for a typical electron temperature $k_{B}T_{e}$=15keV of the CG.  In Fig.~1a, the solid curve is the full calculation with Eq.~(\ref{eq2b-1}) for $P(s,\beta)$, and the dashed curve is the ISO approximation with Eq.~(\ref{eq3c-3}) for $P(s,\beta)$.  In Fig.~1b, the solid curve is the full calculation with Eq.~(\ref{eq2b-2}) for $\tilde{P}_{\rm K}(s,\beta)$, and the dashed curve is the ISO approximation with Eq.~(\ref{eq3c-4}) for $\tilde{P}_{\rm K}(s,\beta)$.  It can be seen that the ISO approximation is an excellent approximation for the SZ effects of the CG.  Our finding confirms the results of the PV paper on the ISO approximation.  The errors of the ISO approximation are 1\% at the peak positions.

\section{Concluding Remarks}

  We studied the covariant formalism for the SZ effects developed by the present authors (NKI).  We obtained the analytic expressions for the redistribution functions $P(s,\beta)$ and $P_{\rm K}(s,\beta)$ of the thermal and kinematical SZ effects for the CG, respectively.  In the NKI formalism, the change of the photon occupation number $\Delta n(x)$ and the spectral intensity function $\Delta I(x)$ can be expressed by the double integral forms.  Thus, their numerical calculations became extremely fast compared with previous numerical calculations.  The numerical programs will be quite useful for the analysis of existing and future observation data of the SZ effects.  They are available upon request from one of the present authors (S. N.).

  We also studied another covariant formalism for the SZ effects recently developed by Poutanen and Vurm.  We showed that the two formalisms were mathematically identical in the Thomson approximation, which is fully valid for the CMB photon energies.  This has concluded that the existing formalisms\cite{wrig79,noza09a,pout10} of the SZ effects for the CG are equivalent in the Thomson approximation.  Thus, the present finding establishes a theoretical foundation for the analysis of the SZ effects for the CG.

  Finally, we calculated the SZ effects in the isotropic photon scattering (ISO) approximation.  The change of the spectral intensity function $\Delta I(x)$ was calculated both in the full expression and in the ISO approximation.  It was shown that the ISO approximation was an excellent approximation compared with the full calculation.  The errors were 1\% for $\Delta I_{t}(x)$ and $\Delta I_{k}(x)$ at the peak positions.

\begin{acknowledgments}
This work is financially supported in part by the Grant-in-Aid of Japanese Ministry of Education, Culture, Sports, Science, and Technology under Contract No. 21540277.  We would like to thank our referee for valuable suggestions.
\end{acknowledgments}


\bibliography{apssamp}

\end{document}